\documentclass[%
 aip,%
 jap,%
 cp,%
 amsmath,amssymb,%
 reprint,%
 groupedaddress,
]{revtex4-1}
\pdfoutput=1
\usepackage{graphicx}
\usepackage{dcolumn}
\usepackage{bm}
\usepackage{float}
\usepackage[mathlines]{lineno}

\begin{document}

\preprint{AIP/123-QED}

\title[Enhanced exchange bias in MnN/CoFe bilayers]{Enhanced exchange bias in MnN/CoFe bilayers after \\high-temperature annealing }

\author{M. Dunz}
 \email{mdunz@physik.uni-bielefeld.de}
 \author{J. Schmalhorst}%
\author{M. Meinert}%
\affiliation{ 
Center for Spinelectronic Materials and Devices, Department of Physics, Bielefeld University, D-33501 Bielefeld, Germany 
}%

\date{\today}

\begin{abstract}
We report an exchange bias of more than $2700$\,Oe at room temperature in MnN/CoFe bilayers after high-temperature annealing. We studied the dependence of exchange bias on the annealing temperature for different MnN thicknesses in detail and found that samples with $t_{\text{MnN}}>32$\,nm show an increase of exchange bias for annealing temperatures higher than $T_{\text{A}}=400\,^{\circ}$C. Maximum exchange bias values exceeding $2000$\,Oe with reasonably small coercive fields around $600$\,Oe are achieved for $t_{\text{MnN}}= 42, 48$\,nm. The median blocking temperature of those systems is determined to be $180\,^{\circ}$C after initial annealing at $T_{\text{A}}=525\,^{\circ}$C. X-ray diffraction measurements and Auger depth profiling show that the large increase of exchange bias after high-temperature annealing is accompanied by strong nitrogen diffusion into the Ta buffer layer of the stacks. 
\end{abstract}

\maketitle

\begin{figure}
\includegraphics[width=0.48\textwidth]{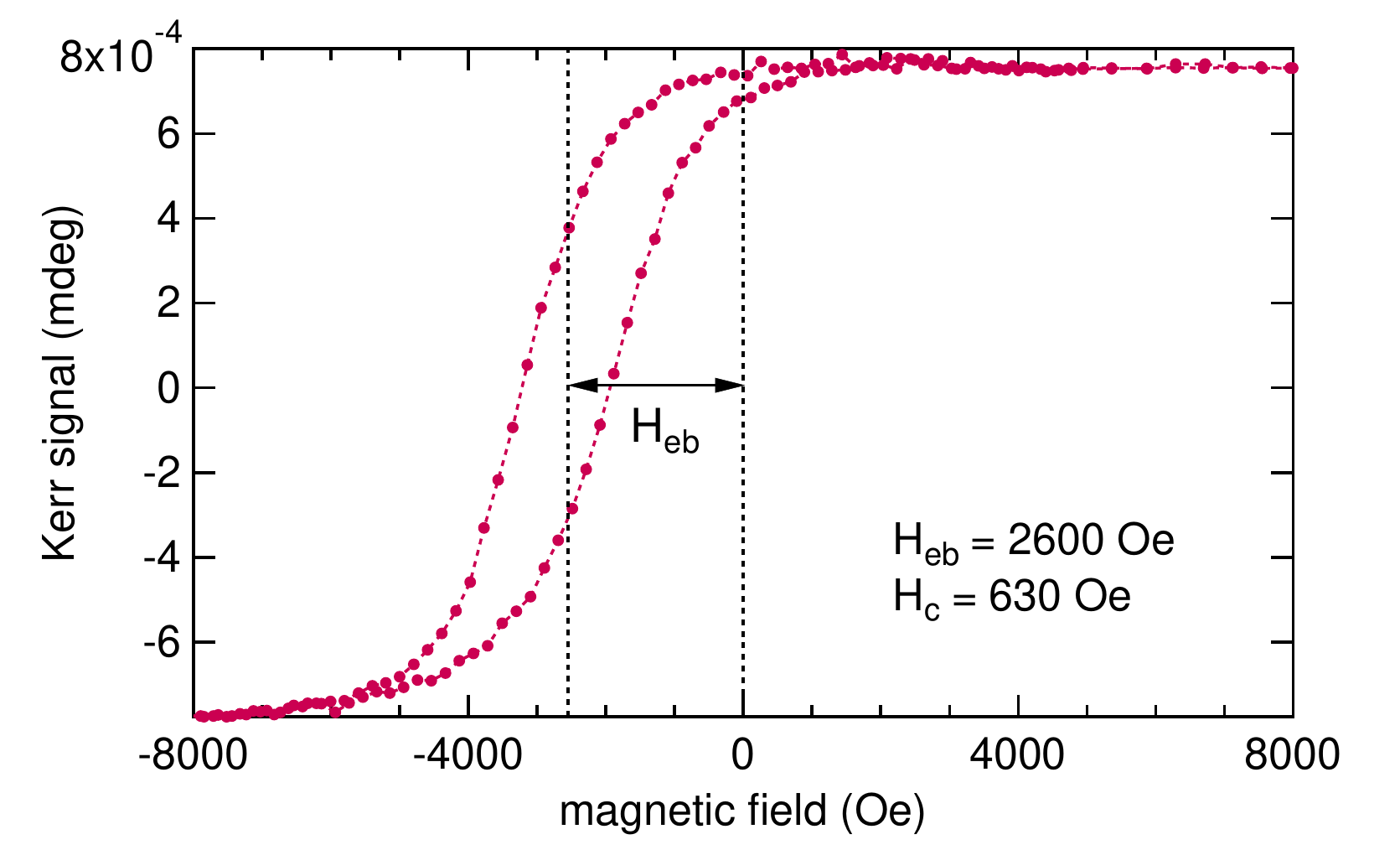}
\caption{Hysteresis loop detected parallel to the field cooling direction for a sample with $t_{\text{CoFe}}=1.6$\,nm and $t_{\text{MnN}}=48$\,nm after annealing at $T_{\text{A}}=500\,^{\circ}$C. The exchange bias field is marked with an arrow.}
\end{figure}
\section{\label{sec:level1}Introduction}
In spinelectronics, the exchange bias effect is used to pin a ferromagnetic
electrode to an antiferromagnetic layer\citep{newmagneticanisotropy, ebias}. This is crucial in GMR or TMR devices to allow for distinct stable resistance states\citep{spintronics}. For several years, the search for new antiferromagnetic materials has been going on in order to find rare-earth free alternatives for commonly used MnIr\citep{mnir} or MnPt\citep{mnpt, mining}. Recently, we found that antiferromagnetic MnN is a promising candidate\citep{meinert, zilske}:  Polycrystalline MnN/CoFe bilayer systems show exchange bias of up to $1800$\,Oe at room temperature. This maximum exchange bias is observed for thicknesses of $t_{\text{MnN}}=30$\,nm and $t_{\text{CoFe}}=1.6$\,nm after annealing and field cooling from $T_{\text{A}}=325\,^{\circ}$C. Significantly higher annealing temperatures lead to a decrease of exchange bias due to nitrogen diffusion. For integration into spinelectronic devices, an increased thermal stability would be desirable.
MnN crystallizes in the $\Theta-$phase of the Mn-N phase diagram\citep{theta}, in a tetragonal variant of the NaCl structure with $a = 4.256$\,\AA{} and $c = 4.189$\,\AA{} at room temperature\citep{Suzuki}. The exact lattice constants depend on the nitrogen content in the lattice. With decreasing nitrogen content, decreasing lattice constants are observed{\citep{Suzuki, Leineweber}. The magnetic moments of MnN are coupled parallel in the $ab$ planes and alternate along the $c$ direction but the spin orientation also depends on the nitrogen content in the lattice\citep{Leineweber, Suzuki2}. The N\'eel temperature of MnN is about $390\,^{\circ}$C\citep{neel}.
\begin{figure*}
\includegraphics[width=0.85\textwidth]{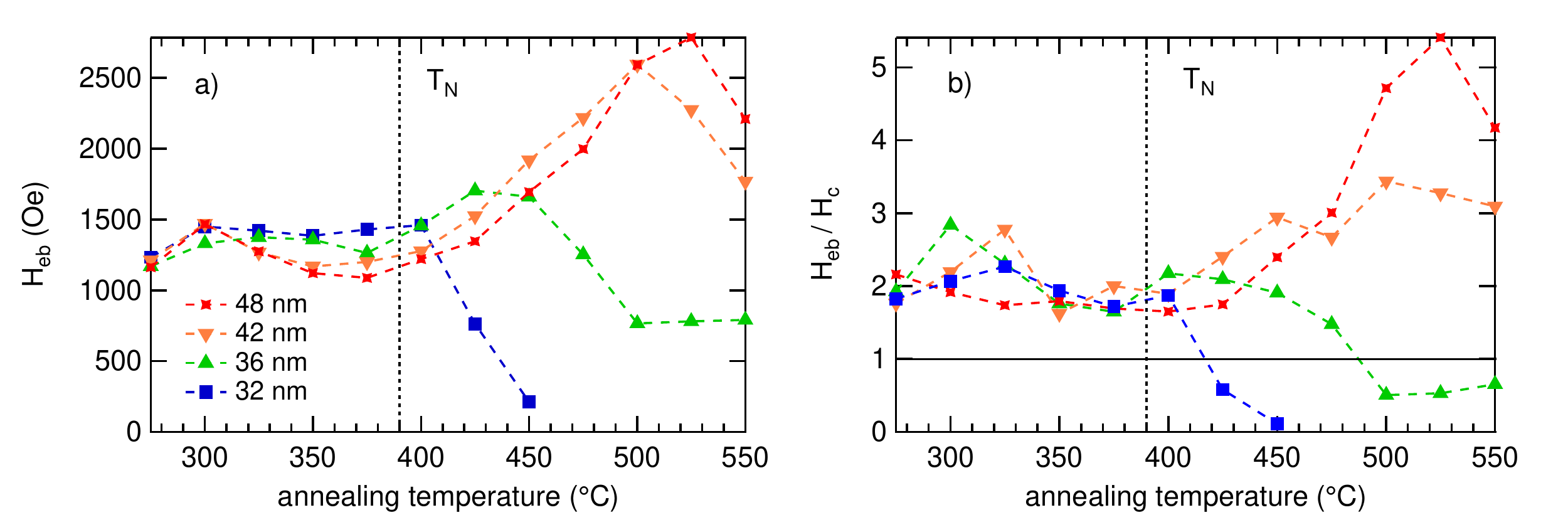}
\caption{Dependence of exchange bias (a) and ratio $H_{\text{eb}}/H_{\text{c}}$ (b) on the annealing temperature measured for samples with different thicknesses of MnN. The crossing of the N\'eel temperature of MnN is indicated with dashed lines.}
\end{figure*}
\section{\label{sec:level1}Experimental}
Ta ($10$\,nm) / MnN ($t_{\text{MnN}}$) / CoFe ($1.6$\,nm) / $\text{TaO}_{\text{x}}$ ($2.5$\,nm) stacks were prepared on thermally oxidized Si substrates via DC magnetron sputtering at room temperature. The MnN film was sputtered reactively from an elemental Mn target with a gas ratio of $50\,\%$ Ar to $50\,\%$ $\text{N}_2$, following the same procedure as described in our previous article\citep{meinert}. Post-annealing for 15 minutes and subsequent field cooling in a magnetic field of $H_{\text{fc}}=6.5$\,kOe parallel to the film plane was performed in a vacuum furnace with pressure below $5\cdot10^{-6}$\,mbar to activate exchange bias. Magnetic characterization of the samples was performed using the longitudinal magneto-optical Kerr effect (MOKE) at room temperature. For annealing series, samples were successively annealed and measurements were taken in between the single steps. Structural characterization was performed via X-ray diffraction measurements with a Philips X'Pert Pro MPD, equipped with a Cu source and Bragg-Brentano optics. To investigate nitrogen diffusion inside the stacks, Auger depth profiling with a scanning Auger microscope PHI660 was used. The samples were continuously rotated during sputtering with a $500$\,eV Ar$^+$ ion beam to achieve optimum depth resolution. The measured Auger intensities are defined as peak to peak heights of the differential spectrum of the different components. Target factor analysis\citep{aes2} was used for separating the different chemical states of nitrogen in $\text{TaN}_{\text{x}}$ and $\text{MnN}_{\text{x}}$.
\section{\label{sec:level1}Results and Discussion}
\begin{figure}
\includegraphics[width=0.5\textwidth]{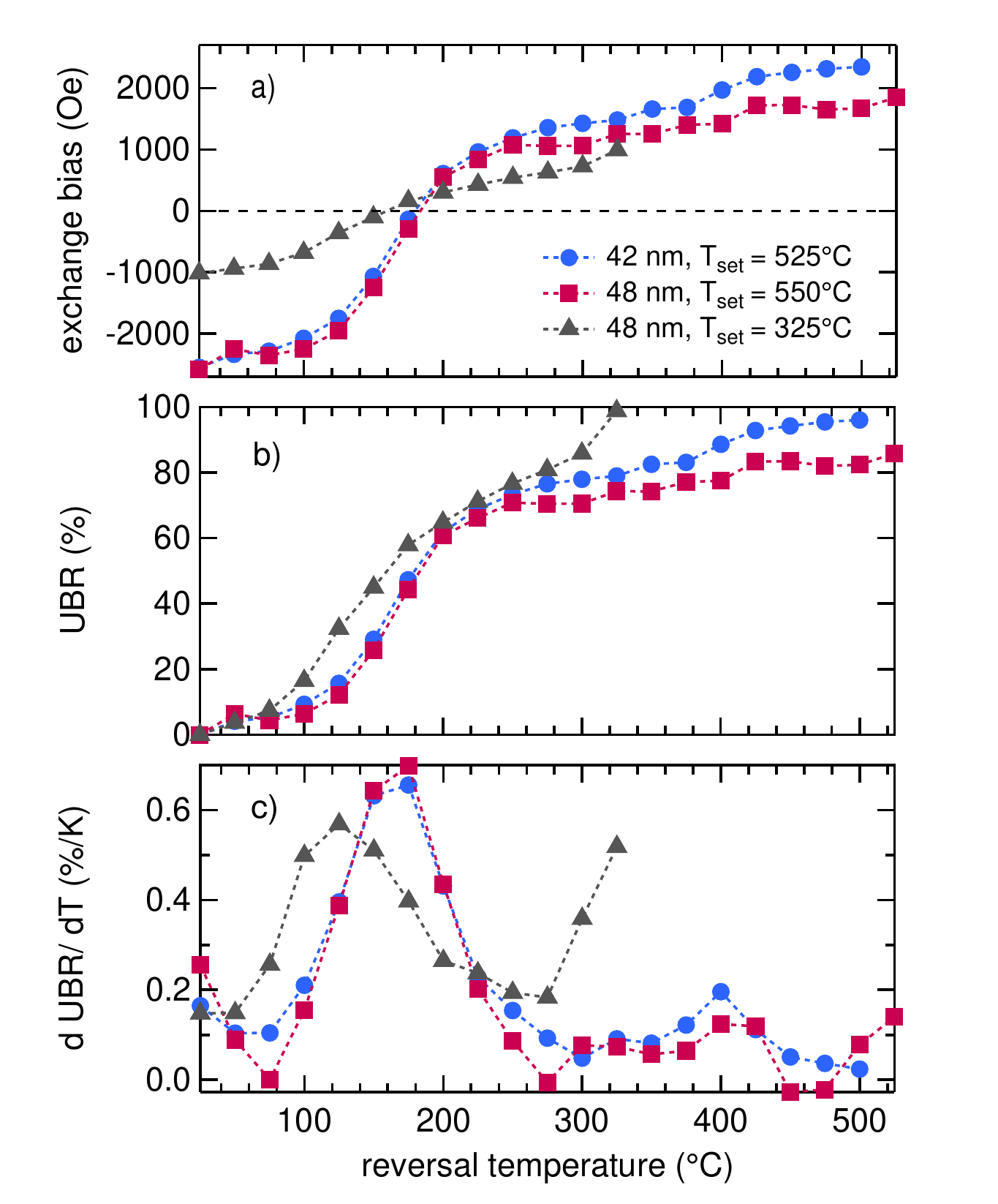}
\caption{Results of the reversed field cooling experiments performed on samples with $t_{\text{MnN}}=42,48$\,nm and $t_{\text{CoFe}}=1.6$\,nm: a) exchange bias, b) unblocked ratio UBR and c) derivative of the unblocked ratio in dependence on the temperature of reversed field cooling.}
\end{figure}
Figure 1 shows an exemplary hysteresis loop detected for a sample with $t_{\text{MnN}}=48$\,nm after annealing at $T_{\text{A}}=500\,^{\circ}$C. Very high exchange bias $H_{\text{eb}}$ accompanied by a reasonably small coercive field $H_{\text{c}}$ can be observed. At zero field, the CoFe layer is almost saturated. We can estimate the maximum effective interfacial exchange energy $J_\mathrm{eff} = t_\mathrm{CoFe} M_\mathrm{CoFe} \mu_0 H_\mathrm{eb}\approx0.76$\, erg/cm$^2$ using the saturation magnetization of $M_{\text{CoFe}}\approx 1700$\, emu/cm$^3$ for our Co$_{70}$Fe$_{30}$ composition\citep{cofe}.\\
In Figure 2a), the detailed dependence of exchange bias on the annealing temperature is shown for different MnN thicknesses. Additionally, the dependence of the ratio of exchange bias and coercive field is displayed in Figure 2b) as it provides information about the possible use in GMR or TMR stacks where $H_{\text{eb}}/H_{\text{c}}>1$ is required. In both graphs, it can clearly be seen that the thermal stability of the samples increases with increasing MnN thickness. With a thickness of $t_{\text{MnN}}=32$\,nm, the exchange bias is stable at a maximum value around $1400$\,Oe up to $T_{\text{A}}=400\,^{\circ}$C before it decreases. In contrast, samples with thicker MnN show an increase of exchange bias after annealing at higher temperatures. This is especially observable for the samples with $t_{\text{MnN}}=42,48$\,nm. Both yield exchange bias values of more than $2500$\,Oe with the maximum of $2785$\,Oe reached after annealing at $T_{\text{A}}=525\,^{\circ}$C for $t_{\text{MnN}}=48$\,nm. The highest ratio of $H_{\text{eb}}/H_{\text{c}}=5.4$ is found for the same parameters.
In our recent article\citep{meinert}, we reported that the thermal stability of MnN crucially depends on the nitrogen content in the MnN lattice. Samples that were prepared with a higher amount of nitrogen ($55\,\%$) during the reactive sputtering showed an increase of exchange bias after annealing at $T_{\text{A}}=400\,^{\circ}$C already for thicknesses of $t_{\text{MnN}}=30$\,nm. As thicker MnN films have a larger nitrogen reservoir, the increase of exchange bias after annealing at high temperatures is now observable for the lower amount of nitrogen ($50\,\%$) during reactive sputtering. Hence, we conclude that the effect of large exchange bias after 
high-temperature annealing is related to the amount of nitrogen in the MnN film.\\
To obtain information about the influence of high-temperature annealing on the blocking temperature distribution, reversed field cooling experiments\citep{blocking} were performed on samples with $t_{\text{MnN}}=42,48$\,nm. The samples were initially field cooled from $T_{\text{set}}=525\,^{\circ}$C and $550\,^{\circ}$C for $t_{\text{MnN}}=42$\,nm and 48\,nm, respectively, and their hysteresis loops were measured to detect $H_{\text{eb}}(\text{RT})$. After that, they were successively field cooled in a reversed field from $T_{\text{rev}}=50, 75,..., 525\,(, 550)\,^{\circ}$C. In Figure 3a), the dependence of exchange bias on the reversal temperature is shown. The zero of this curve marks the median blocking temperature $<T_{\text{B}}>$ of the antiferromagnetic grains that are still blocked at room temperature. For both MnN thicknesses it lies around $180\,^{\circ}$C. The unblocked ratio (UBR), i.e. the area fraction of unblocked grains, is obtained via
\begin{equation}
\text{UBR}(T_{\text{rev}})= 100\%  \cdot \frac {H_{\text{eb}}(\text{RT})-H_{\text{eb}}(T_{\text{rev}})} {2H_{\text{eb}}(\text{RT})},
\end{equation}
representing the cumulative distribution function of the blocking temperature. By taking the derivative of this function, the blocking temperature distribution can be obtained. It is shown in Figure 3c) and has a maximum around $170\,^{\circ}$C. In the course of our previous investigations\citep{meinert} we already performed field cooling experiments on similar samples with $t_{\text{MnN}}=48$\,nm, but initially field cooled them from a lower temperature $T_{\text{set}}=325\,^{\circ}$C. As additionally shown in Figure 3, they yield a median $<T_{\text{B}}>$ of $160\,^{\circ}$C with a maximum of the corresponding blocking temperature distribution around $125\,^{\circ}$C. Comparing this to our new results, the high-temperature annealing has a positive influence on the thermal stability of the MnN/CoFe system and seems to increase either the grain volume or the magnetocrystalline anisotropy energy.\\
In order to identify what causes the giant increase of exchange bias after high-temperature annealing, changes in the crystal structure were investigated for a sample with $t_{\text{MnN}}=42$\,nm. X-ray diffraction scans were detected between each annealing step. In Figure 4, the corresponding diffraction patterns directly after preparation and after annealing at $T_{\text{A}}=325\,^{\circ}$C and $T_{\text{A}}=525\,^{\circ}$C are shown, confirming a growth in (001) direction. After annealing, the (002) and (004) peaks of MnN are shifted towards higher angles, indicating smaller lattice constants due to nitrogen diffusion. Next to that, the peak intensities increase and the peaks become narrower. This indicates growth of the crystal grains and a relaxation of strain. Both can have a beneficial influence on the exchange bias. Larger crystal grains are also in line with the observed enhancement of the blocking temperature\citep{grady}. After annealing at $T_{\text{A}}=525\,^{\circ}$C, an additional peak around $34^{\circ}$ arises that is not related to any phase of Mn-N. Most likely it is attributed to the formation of $\text{TaN}_{\text{x}}$ in the Ta buffer layer caused by the nitrogen diffusion. The inset in Figure 4 displays the detailed evolution of the lattice constant $c$ in dependence on the annealing temperature. Starting at $c=4.23$\,\AA, it decreases monotonously with increasing annealing temperature up to $T_{\text{A}}=475\,^{\circ}$C where it saturates around $c=4.188$\,\AA, close to the value that was determined by Suzuki et al.\citep{Suzuki}.\\
To verify the temperature dependent nitrogen diffusion that is suggested by the X-ray diffraction results, Auger electron depth profiling was performed on a stack with $t_{\text{MnN}}=48$\,nm. Measurements were taken directly after preparation and after annealing at $T_{\text{A}}=325\,^{\circ}$C and $T_{\text{A}}=550\,^{\circ}$C. Figure 5a) shows the nitrogen concentration detected in the Mn and the Ta buffer layer for those three settings. Obviously, the nitrogen concentration in the Ta layer increases with each annealing process. After annealing at $T_{\text{A}}=550\,^{\circ}$C, an almost homogeneous concentration of nitrogen can be found in the whole Ta film. In Figure 5b), the evolution of the Auger peak of nitrogen for different depths is displayed after annealing at $T_{\text{A}}=550\,^{\circ}$C. The color change from blue to pink relates to increasing sputter time/ depth, i.e. the transition from Mn to Ta. A chemical shift of the nitrogen Auger peak is clearly visible. This can be observed in a similar extent after annealing at $T_{\text{A}}=325\,^{\circ}$C. What also strikes is that the nitrogen concentration inside the Mn layer does not decrease with increasing nitrogen concentration in the Ta layer. A strong preferential sputtering of nitrogen is known for other  transition-metal nitrides\citep{surface}, which can change the apparent composition of the nitride film. Our results suggest, that this effect is also present for Mn-N and that the same sputter equilibrium ratio of Mn and N is reached for the sample in the as prepared state as well as after annealing. This can mask the expected reduction of the nitrogen concentration in the Mn-N if the nitrogen concentration in the Ta increases. Nonetheless, the depth profiles verify strong nitrogen diffusion caused by the high-temperature annealing. This coincides with the appearance of the additional peak in the XRD scan related to $\text{TaN}_{\text{x}}$ after annealing at high temperatures as well as the decrease of the lattice constant $c$ with increasing annealing temperature. However, this strong diffusion of nitrogen does not worsen but significantly increase the exchange bias.
\begin{figure*}
\includegraphics[width=0.82\textwidth]{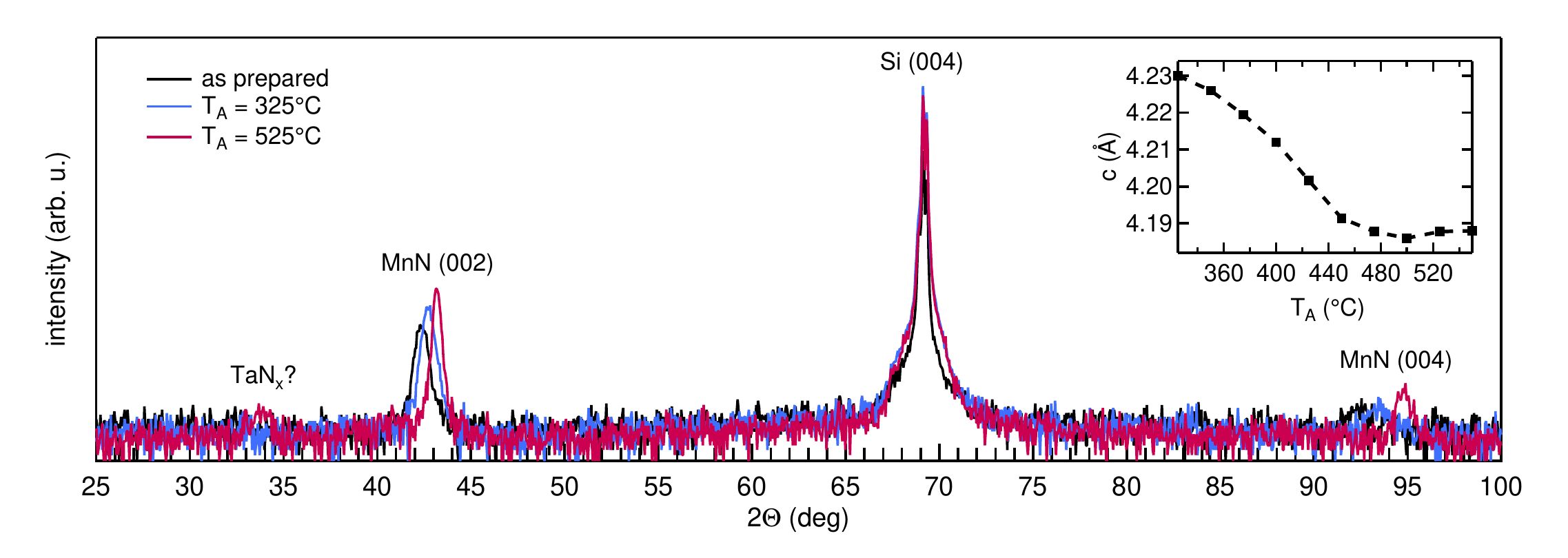}
\caption{X-ray diffraction spectrum of a stack with $t_{\text{MnN}}=42$\,nm before and after annealing at $T_{\text{A}}=325\,^{\circ}$C and $T_{\text{A}}=525\,^{\circ}$C. The inset shows the dependence of the lattice parameter $c$ of MnN on the annealing temperature for the same sample.}
\end{figure*}
\begin{figure*}
\includegraphics[width=0.84\textwidth]{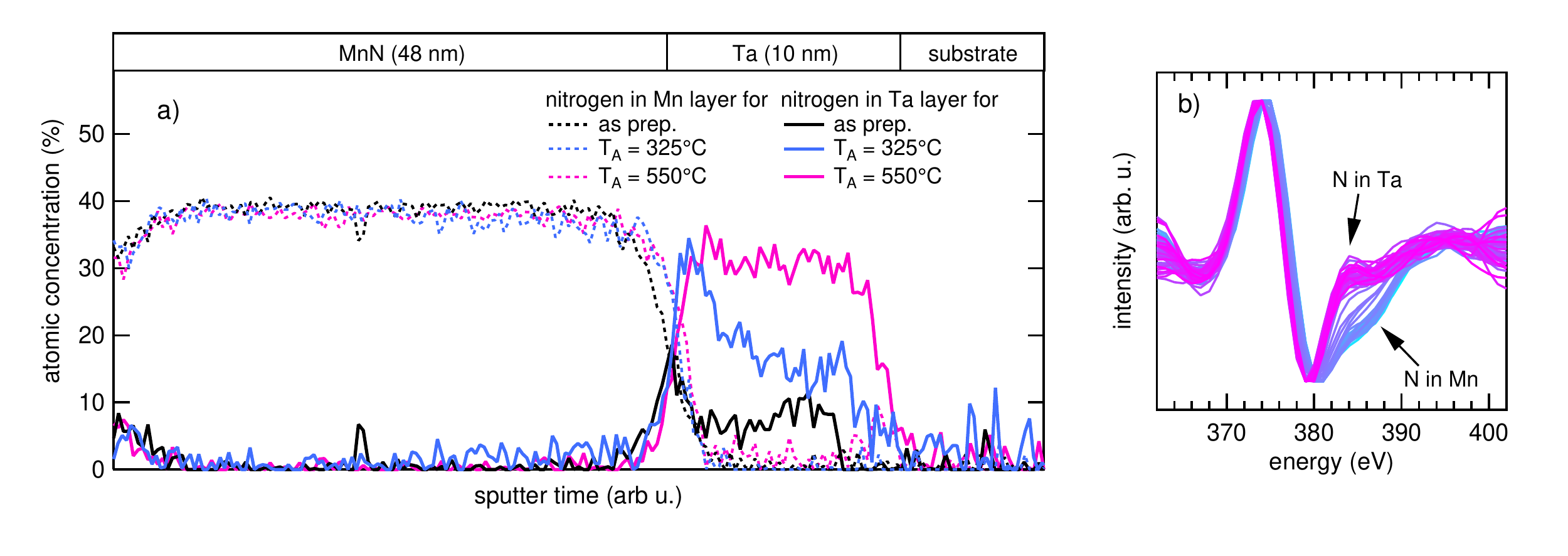}
\caption{a) Depth profile displaying the nitrogen concentration in a stack with $t_{\text{MnN}}=48$\,nm detected with Auger electron spectroscopy directly after preparation and after annealing at $T_{\text{A}}=325\,^{\circ}$C and $T_{\text{A}}=550\,^{\circ}$C. b) Normalized  KLL-Auger transition of nitrogen in differential spectrum for different depths after annealing at $T_{\text{A}}=550\,^{\circ}$C. Color transition from blue to pink corresponds to increasing depth.}
\end{figure*}
\section{\label{sec:level1}Conclusion}
We found that MnN/CoFe bilayers show an enhanced exchange bias after annealing at temperatures $T_{\text{A}} > 400\,^{\circ}$C. Thus, maximum exchange bias of $2785$\,Oe is achieved accompanied by an increased blocking temperature. This behavior is only observable for samples with $t_{\text{MnN}}>32$\,nm. Even though Auger depth profiling confirmed strong nitrogen diffusion into the Ta buffer layer after annealing, thick MnN films seem to have a large nitrogen reservoir that allows for crystallographic or magnetic modifications. They could cause the strong increase of exchange bias and their identification is the subject of further investigations. Furthermore, depositing MnN on a buffer layer that does not bind nitrogen as strongly as Ta would make high-temperature annealing on thinner MnN films possible. The resulting increase of anisotropy could lead to a reduction of the critical thickness of MnN. 
\begin{acknowledgments}
We thank the Ministerium f\"ur Innovation, Wissenschaft und Forschung des Landes Nordrhein-Westfalen (MIWF NRW) for financial support. We further thank G. Reiss for making available laboratory equipment.
\end{acknowledgments}

\begin{thebibliography}{18}%
\makeatletter
\providecommand \@ifxundefined [1]{%
 \@ifx{#1\undefined}
}%
\providecommand \@ifnum [1]{%
 \ifnum #1\expandafter \@firstoftwo
 \else \expandafter \@secondoftwo
 \fi
}%
\providecommand \@ifx [1]{%
 \ifx #1\expandafter \@firstoftwo
 \else \expandafter \@secondoftwo
 \fi
}%
\providecommand \natexlab [1]{#1}%
\providecommand \enquote  [1]{``#1''}%
\providecommand \bibnamefont  [1]{#1}%
\providecommand \bibfnamefont [1]{#1}%
\providecommand \citenamefont [1]{#1}%
\providecommand \href@noop [0]{\@secondoftwo}%
\providecommand \href [0]{\begingroup \@sanitize@url \@href}%
\providecommand \@href[1]{\@@startlink{#1}\@@href}%
\providecommand \@@href[1]{\endgroup#1\@@endlink}%
\providecommand \@sanitize@url [0]{\catcode `\\12\catcode `\$12\catcode
  `\&12\catcode `\#12\catcode `\^12\catcode `\_12\catcode `\%12\relax}%
\providecommand \@@startlink[1]{}%
\providecommand \@@endlink[0]{}%
\providecommand \url  [0]{\begingroup\@sanitize@url \@url }%
\providecommand \@url [1]{\endgroup\@href {#1}{\urlprefix }}%
\providecommand \urlprefix  [0]{URL }%
\providecommand \Eprint [0]{\href }%
\providecommand \doibase [0]{http://dx.doi.org/}%
\providecommand \selectlanguage [0]{\@gobble}%
\providecommand \bibinfo  [0]{\@secondoftwo}%
\providecommand \bibfield  [0]{\@secondoftwo}%
\providecommand \translation [1]{[#1]}%
\providecommand \BibitemOpen [0]{}%
\providecommand \bibitemStop [0]{}%
\providecommand \bibitemNoStop [0]{.\EOS\space}%
\providecommand \EOS [0]{\spacefactor3000\relax}%
\providecommand \BibitemShut  [1]{\csname bibitem#1\endcsname}%
\let\auto@bib@innerbib\@empty
\bibitem [{\citenamefont {Meiklejohn}\ and\ \citenamefont
  {Bean}(1957)}]{newmagneticanisotropy}%
  \BibitemOpen
  \bibfield  {author} {\bibinfo {author} {\bibfnamefont {W.~H.}\ \bibnamefont
  {Meiklejohn}}\ and\ \bibinfo {author} {\bibfnamefont {C.~P.}\ \bibnamefont
  {Bean}},\ }\href@noop {} {\bibfield  {journal} {\bibinfo  {journal} {Phys.
  Rev.}\ }\textbf {\bibinfo {volume} {105}},\ \bibinfo {pages} {904} (\bibinfo
  {year} {1957})}\BibitemShut {NoStop}%
\bibitem [{\citenamefont {Nogués}\ and\ \citenamefont
  {Schuller}(1999)}]{ebias}%
  \BibitemOpen
  \bibfield  {author} {\bibinfo {author} {\bibfnamefont {J.}~\bibnamefont
  {Nogués}}\ and\ \bibinfo {author} {\bibfnamefont {I.~K.}\ \bibnamefont
  {Schuller}},\ }\href@noop {} {\bibfield  {journal} {\bibinfo  {journal} {J.
  Magn. Magn. Mater.}\ }\textbf {\bibinfo {volume} {192}},\ \bibinfo {pages}
  {203} (\bibinfo {year} {1999})}\BibitemShut {NoStop}%
\bibitem [{\citenamefont {Chappert}, \citenamefont {Fert},\ and\ \citenamefont
  {van Dau}(2007)}]{spintronics}%
  \BibitemOpen
  \bibfield  {author} {\bibinfo {author} {\bibfnamefont {C.}~\bibnamefont
  {Chappert}}, \bibinfo {author} {\bibfnamefont {A.}~\bibnamefont {Fert}}, \
  and\ \bibinfo {author} {\bibfnamefont {N.}~\bibnamefont {van Dau}},\
  }\href@noop {} {\bibfield  {journal} {\bibinfo  {journal} {Nature Mater.}\
  }\textbf {\bibinfo {volume} {6}},\ \bibinfo {pages} {813} (\bibinfo {year}
  {2007})}\BibitemShut {NoStop}%
\bibitem [{\citenamefont {Fuke}\ \emph {et~al.}(1997)\citenamefont {Fuke},
  \citenamefont {Saito}, \citenamefont {Kamiguchi}, \citenamefont {Iwasaki},\
  and\ \citenamefont {Sahashi}}]{mnir}%
  \BibitemOpen
  \bibfield  {author} {\bibinfo {author} {\bibfnamefont {H.~N.}\ \bibnamefont
  {Fuke}}, \bibinfo {author} {\bibfnamefont {K.}~\bibnamefont {Saito}},
  \bibinfo {author} {\bibfnamefont {Y.}~\bibnamefont {Kamiguchi}}, \bibinfo
  {author} {\bibfnamefont {H.}~\bibnamefont {Iwasaki}}, \ and\ \bibinfo
  {author} {\bibfnamefont {M.}~\bibnamefont {Sahashi}},\ }\href@noop {}
  {\bibfield  {journal} {\bibinfo  {journal} {J. Appl. Phys.}\ }\textbf
  {\bibinfo {volume} {81}},\ \bibinfo {pages} {4004} (\bibinfo {year}
  {1997})}\BibitemShut {NoStop}%
\bibitem [{\citenamefont {Saito}\ \emph {et~al.}(1997)\citenamefont {Saito},
  \citenamefont {Hasegawa}, \citenamefont {Watanabe}, \citenamefont {Kakihara},
  \citenamefont {Sato}, \citenamefont {Seki}, \citenamefont {Nakazawa},
  \citenamefont {Makino},\ and\ \citenamefont {Kuriyama}}]{mnpt}%
  \BibitemOpen
  \bibfield  {author} {\bibinfo {author} {\bibfnamefont {M.}~\bibnamefont
  {Saito}}, \bibinfo {author} {\bibfnamefont {N.}~\bibnamefont {Hasegawa}},
  \bibinfo {author} {\bibfnamefont {T.}~\bibnamefont {Watanabe}}, \bibinfo
  {author} {\bibfnamefont {Y.}~\bibnamefont {Kakihara}}, \bibinfo {author}
  {\bibfnamefont {K.}~\bibnamefont {Sato}}, \bibinfo {author} {\bibfnamefont
  {H.}~\bibnamefont {Seki}}, \bibinfo {author} {\bibfnamefont {Y.}~\bibnamefont
  {Nakazawa}}, \bibinfo {author} {\bibfnamefont {A.}~\bibnamefont {Makino}}, \
  and\ \bibinfo {author} {\bibfnamefont {T.}~\bibnamefont {Kuriyama}},\ }in\
  \href@noop {} {\emph {\bibinfo {booktitle} {Magnetics Conference, 1997.
  Digests of INTERMAG '97, IEEE International}}}\ (\bibinfo {year} {New
  Orleans, 1997})\BibitemShut {NoStop}%
\bibitem [{\citenamefont {Glaister}\ and\ \citenamefont {Mudd}(2010)}]{mining}%
  \BibitemOpen
  \bibfield  {author} {\bibinfo {author} {\bibfnamefont {B.~J.}\ \bibnamefont
  {Glaister}}\ and\ \bibinfo {author} {\bibfnamefont {G.~M.}\ \bibnamefont
  {Mudd}},\ }\href@noop {} {\bibfield  {journal} {\bibinfo  {journal} {Miner.
  Eng.}\ }\textbf {\bibinfo {volume} {23}},\ \bibinfo {pages} {438} (\bibinfo
  {year} {2010})}\BibitemShut {NoStop}%
\bibitem [{\citenamefont {Meinert}\ \emph {et~al.}(2015)\citenamefont
  {Meinert}, \citenamefont {B\"uker}, \citenamefont {Graulich},\ and\
  \citenamefont {Dunz}}]{meinert}%
  \BibitemOpen
  \bibfield  {author} {\bibinfo {author} {\bibfnamefont {M.}~\bibnamefont
  {Meinert}}, \bibinfo {author} {\bibfnamefont {B.}~\bibnamefont {B\"uker}},
  \bibinfo {author} {\bibfnamefont {D.}~\bibnamefont {Graulich}}, \ and\
  \bibinfo {author} {\bibfnamefont {M.}~\bibnamefont {Dunz}},\ }\href@noop {}
  {\bibfield  {journal} {\bibinfo  {journal} {Phys. Rev. B}\ }\textbf {\bibinfo
  {volume} {92}},\ \bibinfo {pages} {144408} (\bibinfo {year}
  {2015})}\BibitemShut {NoStop}%
\bibitem [{\citenamefont {Zilske}\ \emph {et~al.}(2017)\citenamefont {Zilske},
  \citenamefont {Graulich}, \citenamefont {Dunz},\ and\ \citenamefont
  {Meinert}}]{zilske}%
  \BibitemOpen
  \bibfield  {author} {\bibinfo {author} {\bibfnamefont {P.}~\bibnamefont
  {Zilske}}, \bibinfo {author} {\bibfnamefont {D.}~\bibnamefont {Graulich}},
  \bibinfo {author} {\bibfnamefont {M.}~\bibnamefont {Dunz}}, \ and\ \bibinfo
  {author} {\bibfnamefont {M.}~\bibnamefont {Meinert}},\ }\href@noop {}
  {\bibfield  {journal} {\bibinfo  {journal} {Appl. Phys. Lett.}\ }\textbf
  {\bibinfo {volume} {110}},\ \bibinfo {pages} {192402} (\bibinfo {year}
  {2017})}\BibitemShut {NoStop}%
\bibitem [{\citenamefont {Gokcen}(1990)}]{theta}%
  \BibitemOpen
  \bibfield  {author} {\bibinfo {author} {\bibfnamefont {N.~A.}\ \bibnamefont
  {Gokcen}},\ }\href@noop {} {\bibfield  {journal} {\bibinfo  {journal} {Bull.
  of Alloy Phase Diagr.}\ }\textbf {\bibinfo {volume} {11}},\ \bibinfo {pages}
  {33} (\bibinfo {year} {1990})}\BibitemShut {NoStop}%
\bibitem [{\citenamefont {Suzuki}\ \emph {et~al.}(2000)\citenamefont {Suzuki},
  \citenamefont {Kaneko}, \citenamefont {Yoshida}, \citenamefont {Obi},
  \citenamefont {Fujimori},\ and\ \citenamefont {Morita}}]{Suzuki}%
  \BibitemOpen
  \bibfield  {author} {\bibinfo {author} {\bibfnamefont {K.}~\bibnamefont
  {Suzuki}}, \bibinfo {author} {\bibfnamefont {T.}~\bibnamefont {Kaneko}},
  \bibinfo {author} {\bibfnamefont {H.}~\bibnamefont {Yoshida}}, \bibinfo
  {author} {\bibfnamefont {Y.}~\bibnamefont {Obi}}, \bibinfo {author}
  {\bibfnamefont {H.}~\bibnamefont {Fujimori}}, \ and\ \bibinfo {author}
  {\bibfnamefont {H.}~\bibnamefont {Morita}},\ }\href@noop {} {\bibfield
  {journal} {\bibinfo  {journal} {J. Alloys Compd.}\ }\textbf {\bibinfo
  {volume} {306}},\ \bibinfo {pages} {66} (\bibinfo {year} {2000})}\BibitemShut
  {NoStop}%
\bibitem [{\citenamefont {A.~Leineweber}\ and\ \citenamefont
  {Kockelmann}(2000)}]{Leineweber}%
  \BibitemOpen
  \bibfield  {author} {\bibinfo {author} {\bibfnamefont {H.~J.}\ \bibnamefont
  {A.~Leineweber}, \bibfnamefont {R.~Niewab}}\ and\ \bibinfo {author}
  {\bibfnamefont {W.}~\bibnamefont {Kockelmann}},\ }\href@noop {} {\bibfield
  {journal} {\bibinfo  {journal} {J. Mater. Chem.}\ }\textbf {\bibinfo {volume}
  {10}},\ \bibinfo {pages} {2827} (\bibinfo {year} {2000})}\BibitemShut
  {NoStop}%
\bibitem [{\citenamefont {Suzuki}\ \emph {et~al.}(2001)\citenamefont {Suzuki},
  \citenamefont {Yamaguchi}, \citenamefont {Kaneko}, \citenamefont {Yoshida},
  \citenamefont {Obi}, \citenamefont {Fujimori},\ and\ \citenamefont
  {Morita}}]{Suzuki2}%
  \BibitemOpen
  \bibfield  {author} {\bibinfo {author} {\bibfnamefont {K.}~\bibnamefont
  {Suzuki}}, \bibinfo {author} {\bibfnamefont {Y.}~\bibnamefont {Yamaguchi}},
  \bibinfo {author} {\bibfnamefont {T.}~\bibnamefont {Kaneko}}, \bibinfo
  {author} {\bibfnamefont {H.}~\bibnamefont {Yoshida}}, \bibinfo {author}
  {\bibfnamefont {Y.}~\bibnamefont {Obi}}, \bibinfo {author} {\bibfnamefont
  {H.}~\bibnamefont {Fujimori}}, \ and\ \bibinfo {author} {\bibfnamefont
  {H.}~\bibnamefont {Morita}},\ }\href@noop {} {\bibfield  {journal} {\bibinfo
  {journal} {J. Phys. Soc. Jpn.}\ }\textbf {\bibinfo {volume} {70}},\ \bibinfo
  {pages} {1084} (\bibinfo {year} {2001})}\BibitemShut {NoStop}%
\bibitem [{\citenamefont {Tabuchi}, \citenamefont {Takahashi},\ and\
  \citenamefont {Kanamaru}(1994)}]{neel}%
  \BibitemOpen
  \bibfield  {author} {\bibinfo {author} {\bibfnamefont {M.}~\bibnamefont
  {Tabuchi}}, \bibinfo {author} {\bibfnamefont {M.}~\bibnamefont {Takahashi}},
  \ and\ \bibinfo {author} {\bibfnamefont {F.}~\bibnamefont {Kanamaru}},\
  }\href@noop {} {\bibfield  {journal} {\bibinfo  {journal} {J. Alloys Compd.}\
  }\textbf {\bibinfo {volume} {210}},\ \bibinfo {pages} {143} (\bibinfo {year}
  {1994})}\BibitemShut {NoStop}%
\bibitem [{\citenamefont {Malinowski}(1991)}]{aes2}%
  \BibitemOpen
  \bibfield  {author} {\bibinfo {author} {\bibfnamefont {E.}~\bibnamefont
  {Malinowski}},\ }\href@noop {} {\emph {\bibinfo {title} {Factor Analysis in
  Chemistry}}}\ (\bibinfo  {publisher} {Wiley, New York},\ \bibinfo {year}
  {1991})\BibitemShut {NoStop}%
\bibitem [{\citenamefont {Reck}\ and\ \citenamefont {Fry}(1969)}]{cofe}%
  \BibitemOpen
  \bibfield  {author} {\bibinfo {author} {\bibfnamefont {R.~A.}\ \bibnamefont
  {Reck}}\ and\ \bibinfo {author} {\bibfnamefont {D.~L.}\ \bibnamefont {Fry}},\
  }\href@noop {} {\bibfield  {journal} {\bibinfo  {journal} {Phys. Rev.}\
  }\textbf {\bibinfo {volume} {184}},\ \bibinfo {pages} {492} (\bibinfo {year}
  {1969})}\BibitemShut {NoStop}%
\bibitem [{\citenamefont {Nozières}\ \emph {et~al.}(2000)\citenamefont
  {Nozières}, \citenamefont {Jaren}, \citenamefont {Zhang}, \citenamefont
  {Zeltser}, \citenamefont {Pentek},\ and\ \citenamefont
  {Speriosu}}]{blocking}%
  \BibitemOpen
  \bibfield  {author} {\bibinfo {author} {\bibfnamefont {J.~P.}\ \bibnamefont
  {Nozières}}, \bibinfo {author} {\bibfnamefont {S.}~\bibnamefont {Jaren}},
  \bibinfo {author} {\bibfnamefont {Y.~B.}\ \bibnamefont {Zhang}}, \bibinfo
  {author} {\bibfnamefont {A.}~\bibnamefont {Zeltser}}, \bibinfo {author}
  {\bibfnamefont {K.}~\bibnamefont {Pentek}}, \ and\ \bibinfo {author}
  {\bibfnamefont {V.~S.}\ \bibnamefont {Speriosu}},\ }\href@noop {} {\bibfield
  {journal} {\bibinfo  {journal} {J. Appl. Phys.}\ }\textbf {\bibinfo {volume}
  {87}},\ \bibinfo {pages} {3920} (\bibinfo {year} {2000})}\BibitemShut
  {NoStop}%
\bibitem [{\citenamefont {Gall}\ \emph {et~al.}(2000)\citenamefont {Gall},
  \citenamefont {Haasch}, \citenamefont {Finnegan}, \citenamefont {Lee},
  \citenamefont {Shin}, \citenamefont {Sammann}, \citenamefont {Greene},\ and\
  \citenamefont {Petrov}}]{surface}%
  \BibitemOpen
  \bibfield  {author} {\bibinfo {author} {\bibfnamefont {D.}~\bibnamefont
  {Gall}}, \bibinfo {author} {\bibfnamefont {R.~T.}\ \bibnamefont {Haasch}},
  \bibinfo {author} {\bibfnamefont {N.}~\bibnamefont {Finnegan}}, \bibinfo
  {author} {\bibfnamefont {T.-Y.}\ \bibnamefont {Lee}}, \bibinfo {author}
  {\bibfnamefont {C.-S.}\ \bibnamefont {Shin}}, \bibinfo {author}
  {\bibfnamefont {E.}~\bibnamefont {Sammann}}, \bibinfo {author} {\bibfnamefont
  {J.~E.}\ \bibnamefont {Greene}}, \ and\ \bibinfo {author} {\bibfnamefont
  {I.}~\bibnamefont {Petrov}},\ }\href@noop {} {\bibfield  {journal} {\bibinfo
  {journal} {Surface Science Spectra}\ }\textbf {\bibinfo {volume} {7}},\
  \bibinfo {pages} {167} (\bibinfo {year} {2000})}\BibitemShut {NoStop}%
\bibitem [{\citenamefont {O'Grady}, \citenamefont {Fernandez-Outon},\ and\
  \citenamefont {{Vallejo Fernandez}}(2010)}]{grady}%
  \BibitemOpen
  \bibfield  {author} {\bibinfo {author} {\bibfnamefont {K.}~\bibnamefont
  {O'Grady}}, \bibinfo {author} {\bibfnamefont {L.}~\bibnamefont
  {Fernandez-Outon}}, \ and\ \bibinfo {author} {\bibfnamefont {G.}~\bibnamefont
  {{Vallejo Fernandez}}},\ }\href@noop {} {\bibfield  {journal} {\bibinfo
  {journal} {J. Magn. Magn. Mater.}\ }\textbf {\bibinfo {volume} {322}},\
  \bibinfo {pages} {883} (\bibinfo {year} {2010})}\BibitemShut {NoStop}%
\end{thebibliography}
%

\end{document}